# Geometric and magnetic properties of Co/ Pd system

S.-J.Oh*, Wookje Kim*, Wondong Kim*, B.-H.Choi*, Jae-Young Kim*, Hoon Koh*, H.-J. Kim*, and J.-H.Park**

*Department of physics, Seoul National University, Seoul 151-742, Korea.

**National Synchrotron Light Source, Brookhaven National Laboratory, Upton, New York 11973-5000, U.S.A

Prof. Se-Jung Oh

Department of physics, Seoul National University, Seoul 151-742, Korea.

Telephone : 82-2-880-6609

Fax : 82-2-884-3002

E-mail : sjoh@plaza.snu.ac.kr

**Abstract**

We measured geometric and magnetic properties of Co films on the Pd(111) surface by x-ray photoelectron diffraction (XPD), x-ray magnetic circular dichroism (MCD) at the Co $L_{2,3}$ edge, and the surface magneto-optical Kerr effect (SMOKE) measurements. Co thin films are found to grow incoherently with fcc island structure on the smooth Pd(111) substrate. Comparison of MCD and SMOKE measurements of Co thin films grown on rough and smooth Pd(111) surfaces suggests that perpendicular remanent magnetization and Co orbital moment are enhanced by the rough interface. Pd capping layer also induces perpendicular orbital moment enhancement. These observations indicate the influence of hybridization between Co $3d$ and Pd $4d$ at the interface on the magnetic anisotropy.

## 1. Introduction

Recently, magnetic thin film systems and multilayer systems have been studied very actively because of their interesting magnetic properties and possible applications for practical devices. In particular, studies about Co/Pd and Co/Pt systems have been performed intensively because they show perpendicular magnetic anisotropy (PMA) and large Kerr rotation at short wavelengths, which make them promising candidates for the high-density magneto-optical storage media. But the physical origin of PMA is not yet completely understood partly because PMA depends on various environments such as crystal structure, interface structure, film thickness, growth condition, and so on.

The orbital moment is considered important in PMA because the microscopic origin of magnetic anisotropy is thought to be originated from the spin-orbit interaction. A simple picture for the microscopic origin of magnetocrystalline anisotropy was proposed by Bruno [1], which is in large part supported by experiment [2]. Usually in 3*d* transition metal thin films, the spin-orbit coupling energy is large compared to the anisotropy energy, so the orbital moment can redirect the total moment. The orbital magnetic moment can be enhanced due to the modification of density of states (DOS). At the surface, *d* bands become narrower and cause larger magnetic moment. The change of crystal fields and hybridization also affect the DOS. There are many reports supporting enhanced orbital moments for several film systems theoretically [3] and experimentally [4-8], although in some cases reduced orbital moments are also measured [5]. The hybridization effect was emphasized in Co/Cu(100) [5] and Co/Pt [7,8] systems, and the lowering of symmetry was considered important in Co/Cu(100) [4].

Experimentally it is well known that magnetic properties in thin films vary according to the specimens, growing method, geometry, and so on. The change of geometric structure can induce electronic structure change, which in turn influences magnetic properties. So the stacking sequence or roughness is important in magnetic properties of thin films. In this work, we will discuss the effect of the Pd substrate surface roughness and the Pd capping layer on the magnetic properties of Co/Pd(111) system. On the effect of surface roughness on PMA, two conflicting influences are expected theoretically. First, it will induce positive dipolar surface anisotropy resulting in the increase of perpendicular magnetic property [9]. On the other hand, it acts to reduce magnetocrystalline anisotropy (MCA) [10]. Defects from interdiffusion also reduce magnetic anisotropy [11]. Hence, depending on the strength of each contribution, PMA may be enhanced or reduced by the surface roughness. Experimentally, this interplay exhibits complex behaviors. For example in Fe/Ag(100), surface anisotropy is observed to vary by annealing which changes the roughness [12]. In Co/Pt(111) system, longitudinal Kerr signal appears in Co films on rough substrate though it does not exist on smooth substrate [13]. Co/Pt multilayers show either the increase or the decrease of PMA depending on the growth method [14]. Also in case of Fe/Au(100) the hysteresis shape changes, as the roughness is varied [15].

In this paper, we present the results of XPD, MCD and SMOKE measurements on Co thin films on the Pd(111) surface as a function of Co thickness. Co/Pd multilayers have perpendicular magnetic anisotropy [16] and show enhancement of magnetic moment [6]. Co/Pd bilayer system also shows perpendicular magnetization up to 9ML Co coverage [17]. Theoretically the orbital moment enhancement is expected for Co/Pd(111) [3] and magnetic moment enhancement is expected at the interface for Pd/Co/Pd(111) [18]. To understand the geometric structure of Co films on Pd(111) surface, and its effect on the magnetic properties, we first measured XPD on Co/Pd(111) system as a function of Co coverage. We then performed SMOKE and MCD measurements at Co 2$p$ edge on the Co film *in situ* grown on either the sputtered Pd (111) surface or the annealed surface to understand the effect of the substrate roughness on the magnetic property. To compare structure dependence with hybridization effect we also studied Pd capped samples Pd/ Co/Pd(111) system.

## 2. Experiment

X-ray photoelectron diffraction experiment was performed at Seoul National University with a home-made surface analysis system with the hemispherical electron energy analyzer manufactured by VG Scientific Instruments in England. Mg $K_\alpha$ (h$\nu$=1253.6 eV) line was used as a photon source, and Co was evaporated in situ from the e-beam evaporation source. Pd(111) substrate was cleaned by repeated sputtering and annealing procedure and its surface was checked by low energy electron diffraction (LEED). The surface contamination was checked by measuring oxygen and carbon 1s spectra. The base pressure of the chamber was better than $1\times10^{-10}$ Torr, and the pressure did not rise above $1\times10^{-9}$ Torr during Co deposition.

Magnetic circular dichroism experiment was performed at U4B beamline in National Synchrotron Light Source. The base pressure of the vacuum chamber was $2\times10^{-10}$ Torr. Co and Pd films were evaporated by heating Co and Pd wires with electron beam, keeping the pressure under $1\times10^{-9}$ Torr. The Co coverage we grew was in 2~24Å range and the film thickness was determined by thickness monitor. Pd capping layer thickness was fixed at 5Å for all bilayer system. Film deposition rate was about 1Å/min for both Co and Pd films, and was done at room temperature. The surface contamination was checked by measuring oxygen absorption spectra. To see the effect of rough substrate, some films were made on Pd substrate sputtered by 2keV $Ar^+$ ion. For MCD measurement, we took the X-ray absorption spectra in opposite remanent magnetic field directions with fixed incident photon polarization. Absorption spectra were obtained by measuring the total electron yield current. The external magnetic field (±500 gauss) was applied by pulsed driven electromagnet in situ. For in-plane magnetization we set the sample plane parallel to the magnetic field and photon incidence angle 45°. For perpendicular magnetization, we rotated the sample plane to the 45° off-normal and photon incidence angle was 0°.

Surface magneto-optical Kerr effect was measured in the same chamber as the XPD measurement. The base pressure, cleaning method, and film evaporation method were similar to those of MCD experiment. The film thickness was measured by thickness monitor and calibrated with the calculations from X-ray photoemission spectroscopy. We used 20W He-Ne laser, and the incidence angle was fixed about 30°. An electromagnet, which can supply 500 gauss, was located outside a quartz tube attached to the chamber, and for SMOKE experiment the sample was moved into the quartz tube by the manipulator. The electromagnet was rotated to change the magnetization direction.

## 3. Result and Discussion

Figure 1 shows the XPD pattern of Co $2p$ core-level photoemission as a function of Co coverage on top of the smooth Pd(111) substrate. We took many XPD patterns by changing polar and azimuthal angles, but we only show a representative azimuthal XPD pattern measured at the polar angle of 35° in this figure. We see that even at very low coverage as 0.3ML, the azimuthal XPD pattern appears which looks the same as the Pd(111) substrate. This implies that Co grows in islands and follows the fcc structure as the Pd substrate, as was proposed previously [19]. LEED pattern measured after Co film evaporation shows more diffused pattern as the film thickness increases and this fact is consistent with the previous result that Co film grows incoherently [17].

As for the magnetic properties of Co/Pd(111) system, our MCD and SMOKE measurements show that Co films on the smooth Pd(111) surface show in-plane remanent magnetization for all thickness more than 4Å while perpendicular magnetization can also be seen only in the narrow region around 4Å. But after Pd capping on top of Co films, the magnetization axis rotates from in-plane to out-of-plane and perpendicular magnetization component appears below 20Å Co thickness. In 10Å~20Å range, the in-plane magnetization component co-exists. This is essentially consistent with previous results [17,20].

Figure 2 shows the X-ray absorption and MCD signal for 4Å and 8Å Co coverage films with both smooth and rough Pd substrate surfaces of Co/Pd(111) system, while Fig. 3 shows the same spectra when Pd capping layer was deposited on top of Co/Pd(111) system. We can notice several remarkable changes for Co films deposited on rough Pd substrate in these figures. First, compared with that from smooth Co/Pd(111) interface, the MCD signal from the rough interface is clearly enhanced. Because the MCD intensity is proportional to the magnitude of magnetization [21], we can conclude that the rough interface induces larger remanent magnetization. Another feature of the rough interface system is that the appearance and enhancement of perpendicular magnetization. In the case of rough surface, perpendicular magnetization appears from 4Å and persists up to 8Å thickness, while only in-plan magnetization is visible for the smooth surface. So we can conclude that the rough interface induces perpendicular magnetization in Co/Pd system.

The Pd capping layer also induces the perpendicular magnetization, as previously reported [20], and can be seen from the comparison between Fig. 2 and Fig. 3. For example, in Fig. 3 with Pd capping all 4 Co/Pd systems show perpendicular magnetization, while only one system in Fig. 2 without Pd capping shows perpendicular magnetization. One might suspect the rough interface between Co film and Pd capping layer may induce the perpendicular magnetization for systems with smooth Pd(111) substrate. Indeed from our XPD results discussed above, we can say that Co film deposited on smooth Pd(111) surface is not smooth. However, the data of Pd capped films show that perpendicular magnetization exists even at relatively thick coverage, so this effect is probably not dominant. Instead the hybridization between Co and Pd is probably dominant origin for perpendicular magnetization. Theoretical MCA calculations show that interface anisotropy is enhanced by hybridization between Co 3$d$ and Pd 4$d$ [22]. Also strong hybridization in Pd/Co/Pd(111) is expected theoretically compared to Co/Pd(111) [18]. This is consistent with our experimental data presented above.

One interesting phenomenon is that in 10~20Å thickness region, the in-plane magnetization co-exists in Pd/Co/Pd(111). Similar phenomenon was observed previously in experiments with Au/Co/Au [10] and Co/Pt(111) [8], and can be understood as the rotation of easy axis. Although there is also a possibility that the film has multi-domain as discussed in Co/Pd multilayers [23], this is not reasonable in our case because the film thickness is not so thick and the perpendicular magnetization disappears rather rapidly as the film thickness increases.

We can calculate the orbital magnetic moment of Co films by the sum rules from our X-ray MCD spectra. We find that for thick coverage around 20Å the orbital moments of the in-plane direction are in the range of 0.13~0.16$\mu_B$, which are close to the known bulk value [24]. The perpendicular orbital magnetic moment values are in the range of

0.2~0.3$\mu_B$ for Pd/Co/Pd(111) samples with 8Å~16Å Co coverage. These values are similar to the previous results obtained for Co/Pd multilayer [6,25] and somewhat larger than the calculated value for Co/Pd(111) bilayer [3]. In the range of 10Å to 20Å, where the easy axis rotates, the orbital moment in the easy direction must be higher. Hence we can conclude that the orbital moment in Co films less than 16Å thick on the Pd substrate is clearly enhanced in comparison with the bulk Co.

One interesting observation from our MCD data is that even the in-plane orbital moment seems to be enhanced when thin Co film is deposited on the smooth Pd substrate showing the in-plane magnetic anisotropy. In fact, the in-plane orbital moment enhancement seems comparable to the perpendicular orbital moment enhancement if we assume the spin moment does not change. This shows that the interface roughness is not the dominant cause of the orbital moment enhancement and the Co-Pd hybridization is probably more important in orbital enhancement. This also implies that the orbital moment anisotropy is not a dominant factor for the perpendicular magnetic anisotropy in this case.

SMOKE experiments performed on Co/Pd systme essentially confirmed the above observations for their magnetic properties. We found that Co/Pd(111) samples with smooth interface show longitudinal hysteresis for 4~30Å of Co thickness, while in those with rough interface the polar hysteresis appears below 10Å. The trend is consistent with the MCD measurements discussed above. Figure 4 and Fig. 5 show selected SMOKE data. In these results, the contribution from Pd substrate is subtracted from the raw data to show Co magnetic property, especially at low coverage. For the data with about 15Å Co coverage, we can see that the SMOKE intensity with the rough interface is slightly large, although both show in-plane magnetization. This again is consistent with MCD results.  The Pd capping induces perpendicular magnetization in both films, as shown in Fig. 5.  Systematic study shows that for Pd capped samples, polar hysteresis shows up below 18Å coverage on either smooth or rough Pd substrate, and large enhancement of Kerr signal is observed.  This means that magnetizaton is larger for Pd capped system as observed in MCD measurements.  The crossover thickness measured by SMOKE is also consistent with MCD data and previous results.

## 4. Conclusion

We have performed XPD, MCD, and SMOKE measurements on Co/Pd(111) system to understand their geometric and magnetic properties.  XPD results show that Co grows incoherently with fcc island structure on the Pd(111) substrate, even at very thin coverage.  The MCD intensity shows Co films on Pd(111) have large remanent magnetization with the rough interface. The rough interface also induces perpendicular magnetization and increase of magnetic anisotropy.  The local structure of magnetic film and the hybridization between Co $3d$ and Pd $4d$ seem to be very important for the perpendicular magnetization anisotropy.

**Acknowledgements**

This work is supported in part by the Basic Future Technology Project. Ministry of Science and Technology in Korea, and by the Korean Science and Engineering Foundation through Grant No. 976-0200-005-2.

## references

[1] P. Bruno, Phys. Rev. B 39 (1989) 865.

[2] D. Weller, J. Stöhr, R. Nakajima, A. Carl, M.G. Samant, C. Chappert, R. Mégy, P. Beauvillain, P. Veillet and G.A. Held, Phys. Rev. Lett 75 (1995) 3752.

[3] J.L. Rodríguez-López, J. Dorantes-Dávila and G.M. Pastor, Phys. Rev. B 57 (1998) 1040.

[4] M. Tischer, O. Hjortstam, D. Arvanitis, J.Hunter Dunn, F. May, K. Barberschke, J. Trygg and J.M. Wills, Phys. Rev. Lett. 75 (1995) 1602.

[5] P. Srivastava, F. Whillhelm, A. Ney, M. Farle, H. Wende, N. Haack, G. Ceballos and K. Barberschke, Phys. Rev. B 58 (1998) 5701.

[6] Y. Wu, J. Stöhr, B.D. Hermsmeier, M.G. Samant and D. Weller, Phys. Rev. Lett. 69 (1992) 2307.

[7] N. Nakajima, T. Koide, T. Shidara, M. Miyauchi, H. Fukutani, A. Fujimori, K. Iio, T. Katayama, M. Nyvlt, Y. Suzuki, Phys. Rev. Lett. 81 (1998) 5229.

[8] J. Thiele, C. Boeglin, K. Hricovoni and F. Chevrier, Phys. Rev. B 53 (1996) R11934.

[9] P. Bruno, J. Appl. Phys. 64 (1988) 3153.

[10] C. Chappert and P. Bruno, J. Appl. Phys. 64 (1988) 5736.

[11] J.M. Macraren and R.H. Victora, J. Appl. Phys. 76 (1994) 6069.

[12] D.M. Schaller, D.E. Bürgler, C.M. Schmidt, F. Meisinger and H.-J. Güntherodt, Phys. Rev. B 59 (1999) 14516.

[13] C.S. Shern, J.S. Tsay, H.Y. Her, Y.E. Wu and R.H. Chen, Surf. Sci. 429 (1999) L497.

[14] P.F. Carcia, Z.G. Li and W.B. Zeper, J. Mag. Mag. Mater. 121 (1993) 452.

[15] Y.-L. He and G.-C. Wang, J. Appl. Phys. 76 (1994) 6446.

[16] P.F. Carcia, A.D. Meinhaldt and A. Suna, Appl. Phys. Lett. 47 (1985) 178.

[17] S.T. Purcell, M.T. Johnson, N.W.E. McGee, J.J. de Vries, W.B. Zeper and W. Hoving, J. Appl. Phys. 73 (1993) 1360.

[18] S.C. Hong, T.H. Rho and J.I. Lee, J. Mag. Mag. Mater. 140-144 (1995) 697.

[19] A. Atrei, G. Rovida, M. Torrini, U. Bardi, M. Gleeson and C.J. Barnes, Surf. Sci. 372 (1997) 91.

[20] B.N. Engel, M.H. Wiedmann, R.A. Van Leeuwen and C.M. Falco, J. Mag. Mag. Mater. 26 (1993) 532.

[21] W.L. O'Brien and B.P. Tonner, Phys. Rev. B 50 (1994) 2963.

[22] D.-S. Wang, R. Wu and A.J. Freeman, Phys. Rev. B 48 (1993) 15886.

[23] S.K. Kim, V.A. Chernov and Y.-M. Koo, J. Mag. Mag. Mater. 170 (1997) L7.

[24] C.T. Chen, Y.U. Idzerda, H.-J. Lin, N.V. Smith, G. Meigs, E. Chaban, G.H. Ho, E. Pellegrin and F. Sette, Phys. Rev. Lett. 75 (1995) 152.

[25] D. Weller, Y. Wu, J. Stöhr, M.G. Samant, B.D. Hermsmeier and C. Chappert, Phys. Rev. B. 49 (1994) 12888.

**Figure Captions**

Fig. 1. Azimuthal XPD pattern measured at the polar angle of 35° for Co/Pd(111).

Fig. 2. X-ray absorption and MCD spectra for 4Å and 8Å Co coverage films with both smooth and rough Pd substrates (a) 4Å smooth surface, in-plane magnetization (b) 4Å rough surface, perpendicular magnetization (c) 8Å smooth surface, in-plane magnetization (d) 8Å rough surface, in-plane magnetization

Fig. 3. X-ray absorption and MCD spectra for 4Å and 8Å Co coverage films with both smooth and rough Pd substrates with Pd capping layer. (a) 4Å smooth surface, perpendicular magnetization (b) 4Å rough surface, perpendicular magnetization (c) 8Å smooth surface, perpendicular magnetization (d) 8Å rough surface, perpendicular magnetization

Fig. 4. The SMOKE signals of Co/Pd(111). (a) 5Å Co (longitudinal), (b) 15Å Co (polar) with smooth interface and (c) 5Å Co (polar), (d) 15Å Co (polar) with rough interface.

Fig. 5. The SMOKE signals of Co/Pd(111) with Pd capping layer. (a) 15Å Co (longitudinal), (b) 23Å Co (polar) with smooth interface and (c) 15Å Co (polar), (d) 22Å Co (polar) with rough interface.

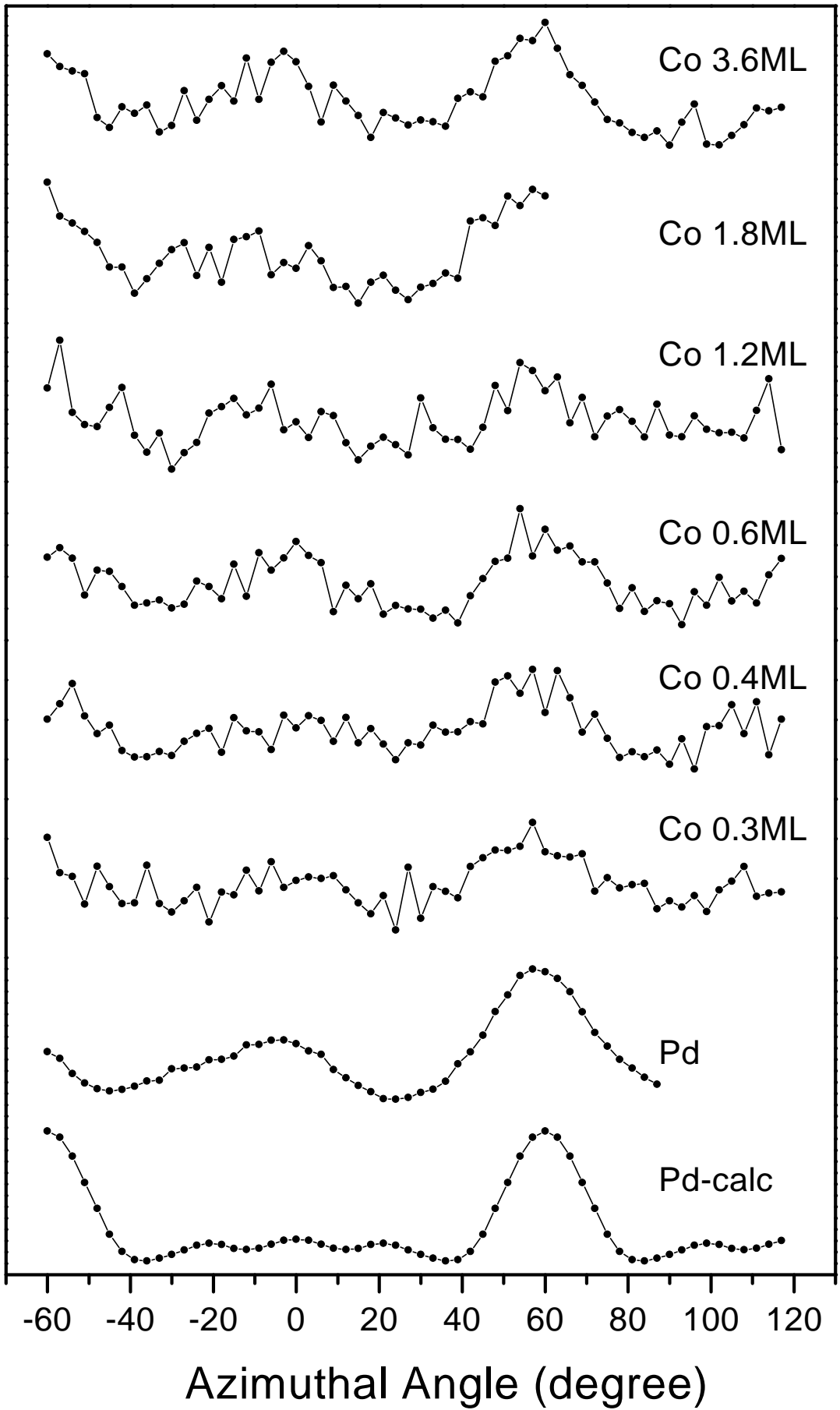
Fig. 1

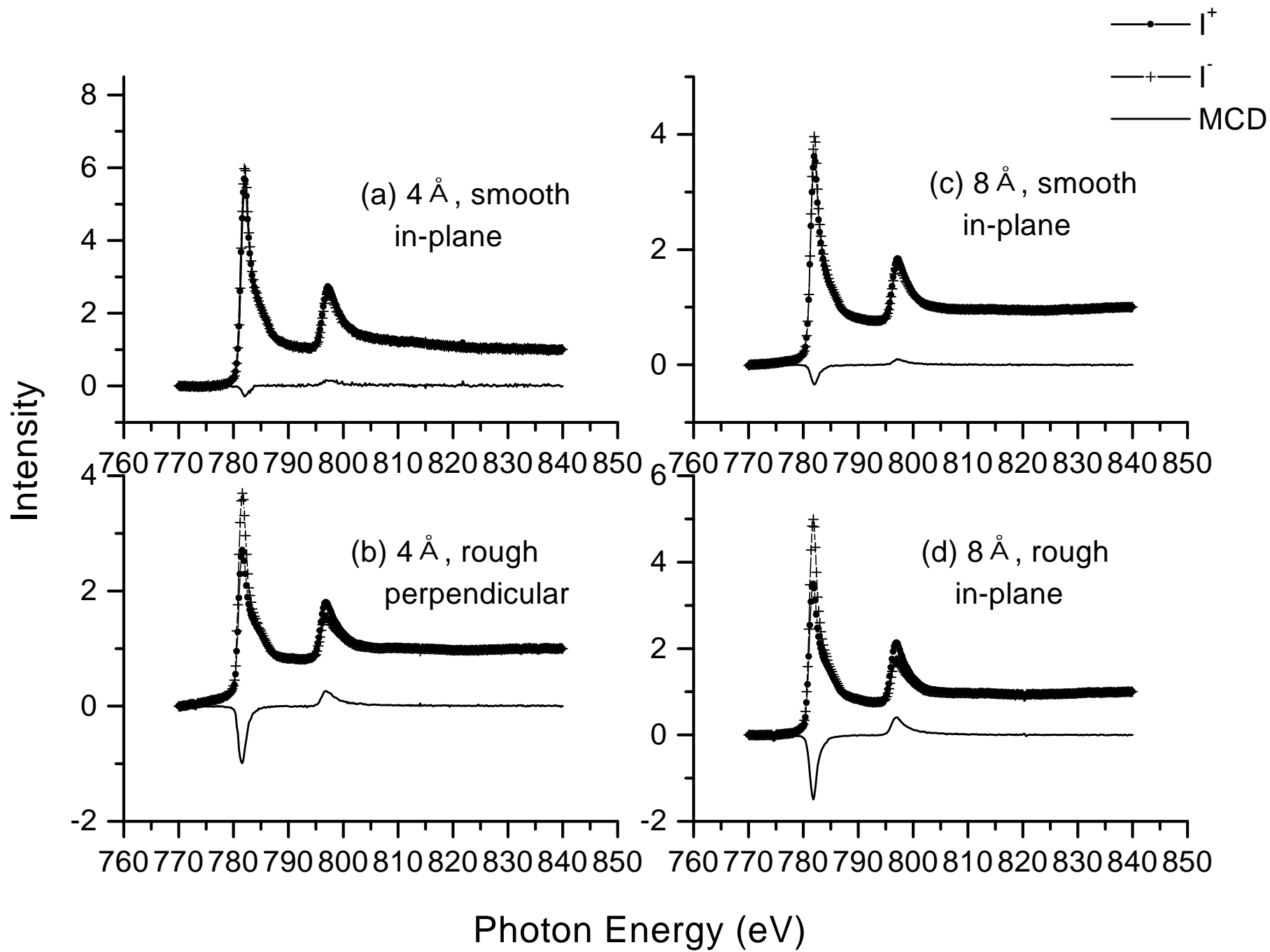

Fig. 2

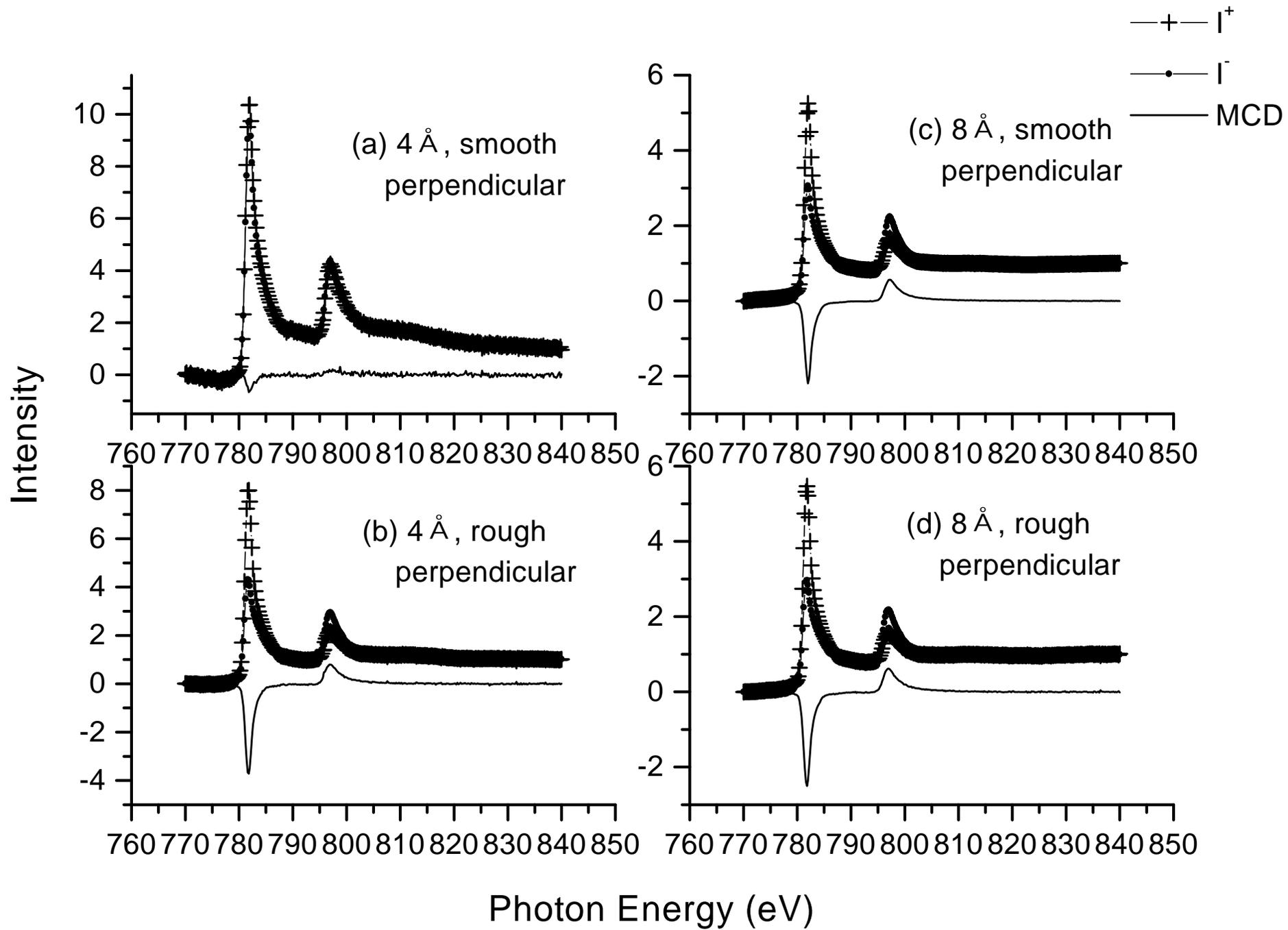

Fig. 3

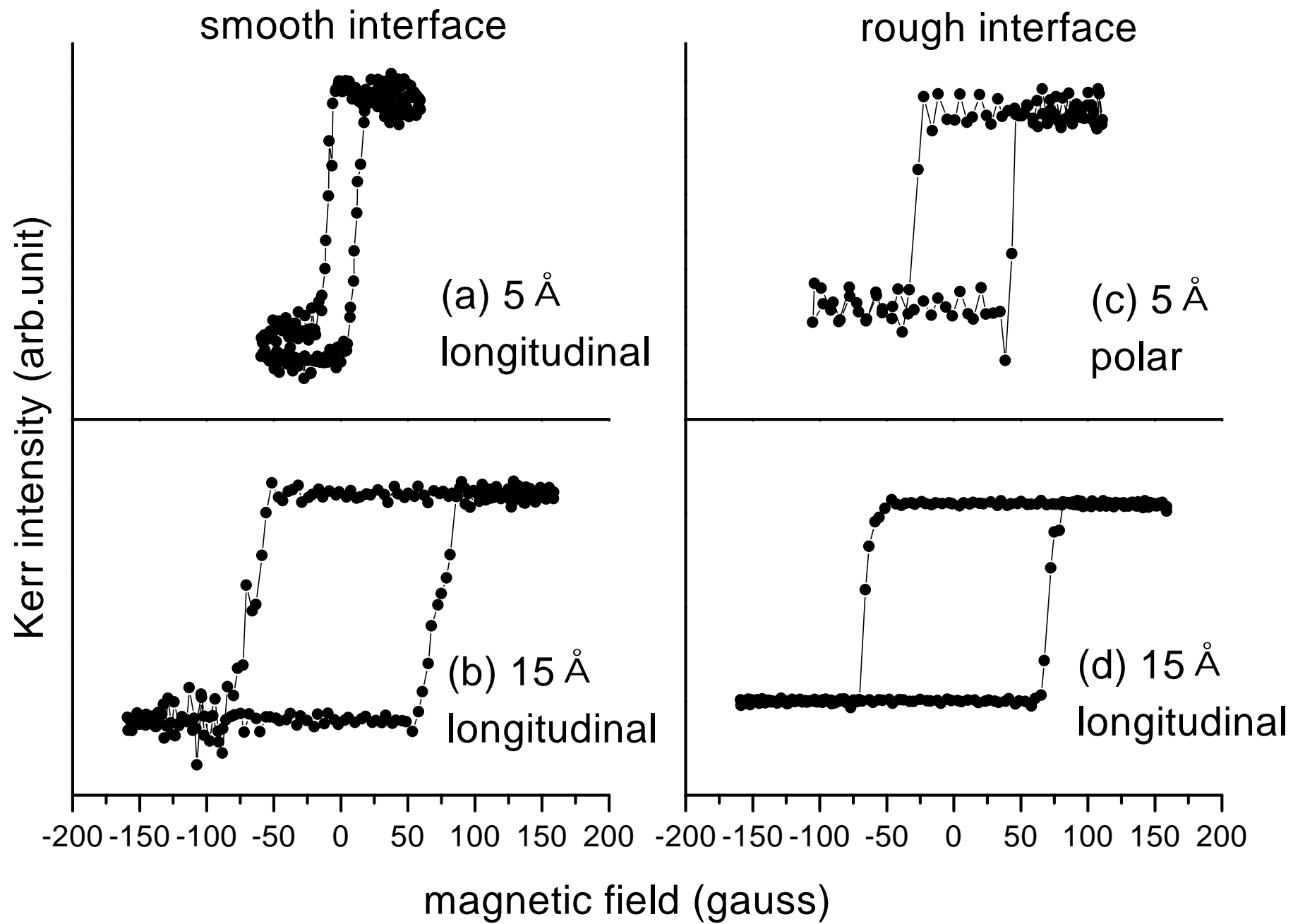

Fig. 4

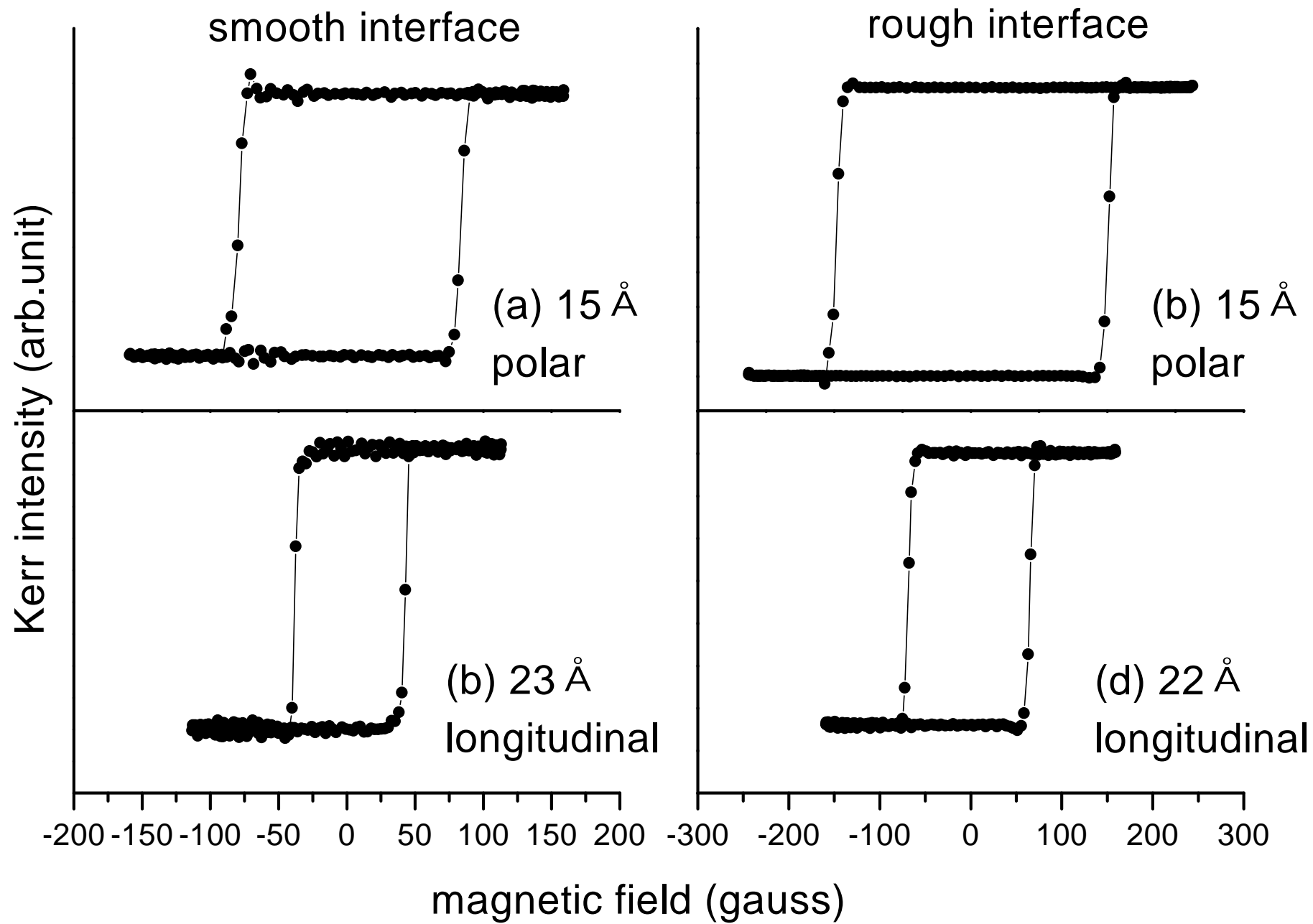

Fig. 5